# High-resolution IR spectroscopy and imaging based on graphene micro-emitters

Kenta Nakagawa,[1,2,‡] Yui Shimura,[1,‡] Yusuke Fukazawa,[1] Ryosuke Nishizaki,[1] Shinichiro Matano,[1] and Hideyuki Maki[1,3,*]

[1]Department of Applied Physics and Physico-Informatics, Keio University, 3-14-1 Hiyoshi, Kohoku-ku, Yokohama, Kanagawa 223-8522, Japan.

[2]Kanagawa Institute of Industrial Science and Technology (KISTEC), 705-1 Shimoimaizumi, Ebina, Kanagawa 243-0435, Japan.

[3]Center for Spintronics Research Network, Keio University, 3-14-1 Hiyoshi, Kohoku-ku, Yokohama, Kanagawa 223-8522, Japan.

[‡]These authors contributed equally to this work.

*E-mail: maki@appi.keio.ac.jp

**Summary paragraph**

IR spectroscopy such as Fourier transform infrared spectroscopy (FTIR) are widely used for the investigation of structure and the quantitative determination of substances in the fields of chemistry, physics, biology, medicine, and astronomy, because the energy of IR absorption corresponds to the energy for each vibrational transition in functional groups within molecules.[1,2,3] Microscopic imaging of FTIR is used for various practical applications, as it enables visualization of the composition distribution and changes in molecular structure without fluorescent labels.[4,5,6,7] However, FTIR microscopy with an objective lens has a diffraction limit causing the low spatial resolution with the order of 10 μm.[3,8,9] Here, we present high-spatial-resolution IR spectroscopy and imaging based on graphene micro-emitters, which have distinct features over conventional IR sources: a planar structure, bright intensity, a small footprint (sub μm$^2$), and high modulation speed of ~100 kHz. We performed IR absorption spectroscopy on a polymer thin film using graphene micro-emitters, realizing high-resolution IR imaging with a spatial resolution of ~2 μm, far higher than that of the conventional FTIR. We show the two-dimensional IR chemical imaging that visualizes the distribution of the chemical information, such as molecular species and functional groups. This technique can open new routes for novel IR imaging and microanalysis in material 32    science, physics, chemistry, biology, and medicine.

**Main text**

The diffraction limit poses a challenge in developing higher-spatial-resolution FTIR. FTIR spectroscopy with an attenuated total reflection crystal is a high-resolution imaging method, which extends the diffraction limit by using a high refractive index crystal.[1,2,3] Atomic force microscope-IR (AFM-IR) spectroscopy[10,11,12,13,14,15,16,17,18] and scattering scanning near-field optical microscopy (s-SNOM)[17,18,19,20,21,22,23,24,25,26,27,28] are methods that extend beyond the diffraction limit of a scanning probe microscope. Although these probe-based methods can realize a high spatial resolution of ~10 nm, an external coherent and strong IR source, such as a tunable IR laser, is required as the strong beam should be focused on the probe tip. However, these tunable IR lasers have very narrow wavelength range, and are also extremely expensive and large, compared with incoherent thermal IR sources, such as incandescent light sources. Moreover, in probe-based IR microscopy, the information for an IR absorption spectra needs to be indirectly extracted from higher-order, modulated, or interferometric signals generated by tip vibration, pulsed laser, or interferometer. The complicated interpretation of the measured signal required for their analyses is in contrast to conventional FTIR.

In this study, we propose novel high-resolution IR absorption spectroscopy and imaging using a graphene micro-emitter as an IR source. Graphene is promising as an IR light source owing to its unique thermal properties, and enables a small footprint, high-speed, on-chip, and bright IR emitter based on blackbody radiation by Joule heating can be fabricated.[29,30,31,32,33,34,35,36,37,38,39,40,41,42,43,44] A graphene emitter does not require an external IR light source (a tunable laser, etc.), so it can enable a small and simple IR spectroscopic technique by the direct modulation of a graphene micro-emitter. In addition, the size of the graphene emitter can be precisely controlled by a micro-fabrication technique in the sub-micrometer range, which is $10^{-6}$ times smaller in area than conventional IR light sources such as incandescent light sources, because graphene can be easily formed by dry-etching processes. Moreover, the graphene micro-emitter can have direct contact with the measured sample, because it has a planar structure with an exposed emitting layer. These features of graphene micro-emitters are advantageous for high-resolution IR imaging. In this study, we fabricated small and bright graphene micro-emitters suitable for IR analysis and successfully demonstrated its application in IR spectroscopy and high-resolution IR imaging.

Blackbody emitters (exactly gray body emitters) with graphene as IR light sources were fabricated on silicon chips. To ensure a small footprint and bright emission, we fabricated the emitter

66  using freely suspended graphene (Fig. 1a).[34] An emitter from unsuspended graphene was also used

67  as a mechanically stable emitter (Extended Data Fig. 1a).[29] For the emitter with suspended graphene,

68  a high temperature can be realized by Joule heating owing to low heat dissipation to a substrate, and

69  the bright blackbody emission can be obtained according to Stefan-Boltzmann law.[34] Details of the

70  fabrication methods of these emitters are described in Methods. Briefly, graphene was transferred on

71  a $SiO_2$ (1500 nm)/Si substrate using a mechanical exfoliation technique. Pd/Cr electrodes were

72  deposited on the graphene as source and drain electrodes. Suspended graphene was formed by $SiO_2$

73  etching using vapor HF. The size of the graphene channel was $5 \times 5$ μm$^2$ for suspended graphene

74  and $0.5 \times 0.5$ μm$^2$ or $10 \times 10$ μm$^2$ for unsuspended graphene. The fabricated chips were mounted on

75  chip carriers and were wire-bonded (Fig. 1a). A near-IR camera image of the thermal emission is

76  observed under a DC bias voltage ($V_{ds}$=1.4 V) in vacuum (Fig. 1b). Bright visible emission can also

77  be observed by a visible camera despite a low bias operation, indicating that the temperature of

78  graphene is remarkably high (Fig. 1b inset).

79  For application in IR analyses of spectroscopy and imaging, we investigated the emission

80  properties of this emitter in mid-IR region. Figure 1c shows the emission spectra (wavelength from

81  2.8 to 4.2 μm under a bias voltage from 1.25 to 1.4 V). Broad emission spectra can be obtained,

82  indicating the blackbody (gray body) radiation generated by Joule heating. The graphene

83  temperatures, obtained by the fitting of Planck's law,[45] are estimated to be 1238, 1340, 1455, and

84  1596 K at $V_{p-p}$=1.25 V, 1.3 V, 1.35 V, and 1.4 V, respectively. These temperatures are significantly

85  higher than the previously reported emitters with unsuspended graphene owing to the suppression of

86  heat dissipation to the substrate[29,30,31,32,33,35,36,37,38,39,41,42,44] and are consistent with the previously

87  reported temperatures in the emitter with the suspended graphene.[34]

88        Previous reports revealed that the emitter with unsuspended graphene has high

89  modulation speed (>1 GHz for single and few layer graphene and several tens MHz for multilayer

90  graphene).[29] We investigated the modulation speed of suspended graphene emitters. Figure 1d shows

91  that the frequency dependences of the light emission intensity were measured by a lock-in amplifier

92  and a spectrum analyzer. In this emitter with suspended graphene, the 3 dB bandwidth of the

93  emission intensity is ~125 kHz, which is slower than that of an unsuspended graphene emitter,

94  because the relaxation time of the graphene temperature is roughly inversely proportional to the heat

95  dissipation to the outside of graphene in the simple thermal model.[29,46,47] However, the bandwidth of

96  125 kHz in our emitter is ~$10^4$ times faster than that in commercially available high-speed IR light

97  sources (~10 Hz) (see Methods and Extended Data Fig. 1c for details). This high-speed modulation

98   can be qualitatively understood by the small heat capacity of graphene owing to an atom-thick

99   substance. These results indicate that the small-footprint, bright, and high-speed blackbody emitters

100  with suspended graphene are suitable for microanalysis system in the IR region. The graphene

101  micro-emitter can realize high spatial resolution of IR spectroscopy and imaging. The high-speed

102  emitter enables the direct modulation of IR light without a light chopper, and this can contribute to

103  downsizing of the analysis system with a lock-in amplifier.

104       In this study, we propose a novel analysis system based on IR spectroscopy and imaging

105  using a graphene-based IR micro-emitter, as shown in Fig. 1e. In this system, the sample is placed

106  over the graphene micro-emitter and is irradiated by IR light. The local IR absorption transmitted

107  through the sample can be directly obtained. Figure 1f shows the schematic picture of the

108  constructed analysis system (see Methods for further details of set-up and measurements). The

109  graphene micro-emitter and the sample mounted to the three-axis stage are placed in the vacuum

110  chamber with an optical window, and the IR light which passes through the sample from the

111  graphene micro-emitter is corrected by an objective lens. The IR light is measured by two different

112  detectors: One is the isolated HgCdTe (MCT) detector, which can directly detect the IR light for

113  high-resolution imaging, and the other is the MCT detector with a monochromator, which can 114   measure the IR spectra for spectroscopy and wavelength-dependent imaging (i.e., chemical imaging).

115   The electrical signal from the MCT detector is measured by a lock-in amplifier, where the IR light

116   from the graphene emitter is directly modulated by applying a rectangular voltage input at 1113 Hz.

117   This is a great feature of the graphene emitter compared with the conventional IR sources, because

118   the high-speed direct modulation can be realized. This direct modulation can also contribute to

119   improved measurement sensitivity as the synchronized IR light to the graphene emission can be

120   selectively measured by a lock-in amplifier without the influence of environmental thermal

121   radiation.

122       We measured absorption spectra of spin-coated polymer thin films on quartz substrate by

123   using a suspended graphene micro-emitter ($5 \times 5$ μm$^2$). Figure 2a shows the absorption spectra of a

124   polymethyl methacrylate (PMMA) film measured by the graphene-based system and conventional

125   FTIR. Absorption peaks corresponding to the overtone of ester $CH_3$ deformation, $CH_3$ (ester methyl)

126   stretching mode, and $CH_3$ (α-methyl) stretching mode of PMMA[48] can be clearly observed at 2841

127   cm$^{-1}$, 2950 cm$^{-1}$, and 2994 cm$^{-1}$, respectively, which is consistent with the FTIR spectra. We also

128   measured the absorption spectra of polystyrene (PS), as shown in Fig. 2b, and absorption peaks

129   corresponding to $CH_2$ symmetric stretching mode, $CH_2$ asymmetric stretching mode, and aromatic

130  CH stretching mode of PS[48] can be observed at 2850 cm$^{-1}$, 2924 cm$^{-1}$, and 3000–3100 cm$^{-1}$,

131  respectively, corresponding to the FTIR spectra.

132  In general, the light intensity detected through a interferometer in FTIR is approximately

133  50% of the energy incident to the spectrometer, whereas the light intensity detected by a

134  monochromator reaches only a few percent of the energy incident to the spectrometer according to

135  so-called Jacquinot advantage.[3,49] Moreover, the graphene microemitter has extremely small

136  footprint, which is approximately 10$^{-6}$ times smaller in area than the ceramic IR sources for FTIR (~

137  mm$^2$). Nevertheless, clear absorption peaks can be observed in the IR spectra measured by the

138  graphene micro-emitter coincide with those of a traditional FTIR. This indicates that graphene

139  micro-emitters are useful for IR spectroscopy, and it is possible to analyze the local molecular

140  structures of substances.

141  We also demonstrated high-resolution IR imaging using an unsuspended graphene

142  micro-emitter with the sub-micrometer size of 0.5 × 0.5 μm$^2$ (the details are described in Methods).

143  Since the specimen is mounted to the three-axis stage, the specimen can be precisely approached to

144  the graphene emitter, and an infrared absorption image can be obtained by scanning the specimen in

145  the *XY* plane.

146 To evaluate the spatial resolution of the infrared absorption imaging, we measured

147 linearly scanned IR absorption of the Ni line and space pattern with the linewidth of 20, 10, 5, and 2

148 μm fabricated by lithography technique on quartz substrate, as shown in Fig. 3a. Figure 3b shows

149 the one-dimensional transmittance along *X* and *Y* axes. The periodical contrast of transmittance,

150 which is owing to the masking of IR light by the Ni line, can be clearly observed for the linewidth of

151 ≥2 μm, a spatial resolution sufficiently higher than conventional FTIR microscopy (see Extended

152 Data Fig. 3).

153 With IR imaging using a graphene micro-emitter, it is expected that the spatial resolution

154 of the imaging is determined by the size of the graphene micro-emitter. To elucidate this, we

155 measured the spatial resolution of this system with shifting the focus of the objective lens. Here, the

156 one-dimensional line scan was performed while shifting the focus of the objective lens in the $\pm z$

157 direction ($|\Delta z|$=50 μm), as shown in Fig. 3c. Surprisingly, no line-profile change is observed with

158 changing the position of the objective lens, indicating that spatial resolution is not dependent on the

159 focus of the objective lens, as shown in Fig. 3d. Here, as the numerical aperture NA of the objective

160 lens used in this study is 0.5, the spot-size diameter *D* of the defocused objective is given by

161 $D=2|\Delta z|\tan(\sin^{-1}\mathrm{NA})\approx 57.7$ μm, which is significantly larger than the width of the line and space.

162  Nevertheless, the clear profile of the line scan with the spatial resolution of ~2 μm can be obtained,

163  independent of the objective focusing. This indicates that spatial resolution can be determined by the

164  size of graphene micro-emitter. The contribution of this size effect to the spatial resolution is

165  maximized on minimizing the gap between the graphene micro-emitter and the sample. As the

166  graphene micro-emitter can be in direct contact with the measured sample, high spatial resolution of

167  IR imaging can be realized by using graphene micro-emitters.

168  To demonstrate high-resolution two-dimensional infrared absorption imaging with a

169  graphene micro-emitter, we measured a 5 μm width '5'-shaped Ni pattern fabricated by

170  photolithography on a quartz substrate as shown in the optical microscope image of Fig. 3e. This

171  patterned specimen was contacted to the graphene micro-emitter and was scanned in the *XY* plane.

172  The two-dimensional image of the transmittance is shown in Fig. 3f. An IR image corresponding to

173  the '5'-shaped pattern can be clearly observed. Notably, the spatial resolution of this IR image by the

174  graphene micro-emitter is significantly higher than by conventional FTIR microscope, as shown in

175  Fig. 3g, in which the image is blurred and the '5' shape cannot be clearly identified.

176  We also demonstrated IR chemical imaging that visualizes the two-dimensional

177  distribution of chemical information, such as molecular species and functional groups using an

178    unsuspended graphene micro-emitter with the size of 10 × 10 μm². Here, we measured the

179    position-dependent IR absorption spectra and two-dimensional IR absorption imaging at the specific

180    wavelength using a monochromator. In this study, the photoresist polymer (ZPN1150) showed broad

181    absorption peaks corresponding to the O-H and C-H stretching modes at 3.0 μm (3333 cm$^{-1}$) and 3.4

182    μm (2941 cm$^{-1}$), respectively (Fig. 4a). A 50 μm width '6'-shaped polymer pattern was fabricated by

183    photolithography on a quartz substrate (Fig. 4b). The IR chemical image of this polymer pattern was

184    measured by scanning the specimen in the *XY* plane and measuring the position-dependent

185    absorption spectra. Figure 4c shows the line scan result of absorption spectra along the black broken

186    line on the polymer pattern in Fig. 4b. Clear absorption can be observed at 3.0 and 3.4 μm owing to

187    O-H and C-H stretching modes on the polymer, respectively. Figure 4d, 4e, and 4f shows

188    two-dimensional absorption imaging at the wavelength of 2.5 μm (no absorption), 3.0 μm (O-H

189    absorption), and 3.4 μm (C-H absorption), respectively. No image can be observed at 2.5 μm (Fig.

190    4d), because there is no absorption at this wavelength as shown in Fig. 4a and 4c. Notably, the slight

191    decrease of the transmittance can be observed at the edge of the polymer pattern due to the light

192    scattering at the edge of the pattern. On the other hand, absorption images, consistent with the

193    polymer pattern in Fig. 4b, can be clearly observed at 3.0 and 3.4 μm, corresponding to O-H and C-H stretching modes, as shown in Fig. 4e and 4f, respectively. These results indicate that the two-dimensional IR chemical images can be obtained by scanning the graphene micro-emitter, and this IR analysis based on the graphene micro-emitter is a novel spatial and spectral analysis.

In conclusion, we have developed and demonstrated high-resolution IR spectroscopy and imaging using graphene micro-emitters. The graphene micro-emitter has the advantages of bright intensity, smaller footprint and diffraction limit, and high modulation speed on silicon chips for this novel IR analysis, in contrast to conventional IR sources. This distinct graphene micro-emitter can realize local IR analysis, whose spatial resolution is determined by the size of the graphene micro-emitter. In addition, the high-resolution IR imaging, including the spatial distribution of the chemical information, such as molecular species and functional groups, can also be obtained by scanning the graphene emitter. As the expected size of the graphene light emitters can be ~10 nm in the future,[50] this analysis method has potential in higher-resolution IR microscopy, which is comparable to probe-based IR microscopy, such as AFM-IR and s-SNOM. Furthermore, this method has distinct advantages from AFM-IR and s-SNOM that this graphene-based IR analysis system can be inexpensively and simply constructed because (i) the external tunable laser, which is ultrabright, large, and expensive, is not required, (ii) the wide wavelength range can be easily measured by using the graphene blackbody emitter compared with the tunable laser, and (iii) the IR spectra can be directly obtained by the simple optical set-up and signal processing, in contrast with the complicated interpretation obtained from the higher-order, modulated, or interferometric signal in AFM-IR and s-SNOM. Hence, this IR spectroscopy and microscopy based on graphene micro-emitters can open the door to nanoscale analytical chemistry by the structure and quantitative analysis for any ultra-small matter, such as nanomaterials, biological samples, difficult synthesis substances, and hazardous substances.

218    **References**

219    1.   Günter Gauglitz and David S. Moore. *Handbook of Spectroscopy: Second, Enlarged Edition*.

220         (WILEY-VCH, 2014).

221    2.   Salzer, R. & Siesler, H. W. *Infrared and Raman Spectroscopic Imaging*. (WILEY-VCH, 2009).

222    3.   Griffiths, P. R. & de Haseth, J. A. *Fourier Transform Infrared Spectrometry (Second Edition)*.

223         (Wiley-Interscience, 2007).

224    4.   Levin, I. W. & Bhargava, R. Fourier Transform Infrared Vibrational Spectroscopic Imaging:

225         Integrating Microscopy and Molecular Recognition. *Annu. Rev. Phys. Chem.* **56**, 429–474 (2005).

226    5.   Sergei G. Kazarian & Chan, K. L. A. Micro- and Macro- Attenuated Total Reflection Fourier

227         Transform Infrared Spectroscopic Imaging. *Appl. Spectrosc.* **64**, 135A-152A (2010).

228    6.   Bhargava, R. Infrared spectroscopic imaging: The next generation. *Appl. Spectrosc.* **66**, 1091–

229         1120 (2012).

230    7.   Bellisola, G. & Sorio, C. Infrared spectroscopy and microscopy in cancer research and diagnosis.

231         *Am. J. Cancer Res.* **2**, 1–21 (2012).

232    8.   Abbe, E. Beiträge zur Theorie des Mikroskops und der mikroskopischen Wahrnehmung. *Arch.*

233         *für Mikroskopische Anat.* **9**, 413–468 (1873).

234    9.    McCutchen, C. W. Superresolution in microscopy and the Abbe resolution limit. *J. Opt. Soc. Am.*

235          **57**, 1190–1192 (1967).

236    10.   Dazzi, A., Prazeres, R., Glotin, F. & Ortega, J. M. Local infrared microspectroscopy with

237          subwavelength spatial resolution with an atomic force microscope tip used as a photothermal

238          sensor. *Opt. Lett.* **30**, 2388–2390 (2005).

239    11.   Dazzi, A., Glotin, F. & Carminati, R. Theory of infrared nanospectroscopy by photothermal

240          induced resonance. *J. Appl. Phys.* **107**, 124519 (2010).

241    12.   Lu, F. & Belkin, M. A. Infrared absorption nano-spectroscopy using sample photoexpansion

242          induced by tunable quantum cascade lasers. *Opt. Express* **19**, 19942–19947 (2011).

243    13.   Lu, F., Jin, M. & Belkin, M. A. Tip-enhanced infrared nanospectroscopy via molecular expansion

244          force detection. *Nat. Photonics* **8**, 307–312 (2014).

245    14.   Ramer, G., Reisenbauer, F., Steindl, B., Tomischko, W. & Lendl, B. Implementation of

246          Resonance Tracking for Assuring Reliability in Resonance Enhanced Photothermal Infrared

247          Spectroscopy and Imaging. *Appl. Spectrosc.* **71**, 2013–2020 (2017).

248    15.   Dazzi, A. *et al.* AFM–IR: Combining Atomic Force Microscopy and Infrared Spectroscopy for

249          Nanoscale Chemical Characterization. *Appl. Spectrosc.* **66**, 1365–1384 (2012).

250    16.  Dazzi, A. & Prater, C. B. AFM-IR: Technology and applications in nanoscale infrared

251        spectroscopy and chemical imaging. *Chem. Rev.* **117**, 5146–5173 (2017).

252    17.  Centrone, A. Infrared Imaging and Spectroscopy Beyond the Diffraction Limit. *Annu. Rev. Anal.*

253        *Chem.* **8**, 101–126 (2015).

254    18.  Xiao, L. & Schultz, Z. D. Spectroscopic Imaging at the Nanoscale: Technologies and Recent

255        Applications. *Anal. Chem.* **90**, 440–458 (2018).

256    19.  Keilmann, F. & Hillenbrand, R. Near-field microscopy by elastic light scattering from a tip.

257        *Philos. Trans. R. Soc. A Math. Phys. Eng. Sci.* **362**, 787–805 (2004).

258    20.  Huth, F., Schnell, M., Wittborn, J., Ocelic, N. & Hillenbrand, R. Infrared-spectroscopic

259        nanoimaging with a thermal source. *Nat. Mater.* **10**, 352–356 (2011).

260    21.  Huth, F. *et al.* Nano-FTIR absorption spectroscopy of molecular fingerprints at 20 nm spatial

261        resolution. *Nano Lett.* **12**, 3973–3978 (2012).

262    22.  Govyadinov, A. A., Amenabar, I., Huth, F., Scott Carney, P. & Hillenbrand, R. Quantitative

263        measurement of local infrared absorption and dielectric function with tip-enhanced near-field

264        microscopy. *J. Phys. Chem. Lett.* **4**, 1526–1531 (2013).

265    23.  Amenabar, I. *et al.* Structural analysis and mapping of individual protein complexes by infrared

266	nanospectroscopy. *Nat. Commun.* **4**, 2890 (2013).

267	24.	Berweger, S. *et al.* Nano-chemical infrared imaging of membrane proteins in lipid bilayers. *J. Am.*

268	*Chem. Soc.* **135**, 18292–18295 (2013).

269	25.	Mastel, S., Govyadinov, A. A., de Oliveira, T. V. A. G., Amenabar, I. & Hillenbrand, R.

270	Nanoscale-resolved chemical identification of thin organic films using infrared near-field

271	spectroscopy and standard Fourier transform infrared references. *Appl. Phys. Lett.* **106**, 023113

272	(2015).

273	26.	Muller, E. A., Pollard, B. & Raschke, M. B. Infrared chemical nano-imaging: Accessing structure,

274	coupling, and dynamics on molecular length scales. *J. Phys. Chem. Lett.* **6**, 1275–1284 (2015).

275	27.	Tranca, D. E. *et al.* High-resolution quantitative determination of dielectric function by using

276	scattering scanning near-field optical microscopy. *Sci. Rep.* **5**, 11876 (2015).

277	28.	Amenabar, I. *et al.* Hyperspectral infrared nanoimaging of organic samples based on Fourier

278	transform infrared nanospectroscopy. *Nat. Commun.* **8**, 14402 (2017).

279	29.	Miyoshi, Y. *et al.* High-speed and on-chip graphene blackbody emitters for optical

280	communications by remote heat transfer. *Nat. Commun.* **9**, 1279 (2018).

281	30.	Freitag, M., Chiu, H. Y., Steiner, M., Perebeinos, V. & Avouris, P. Thermal infrared emission

282         from biased graphene. *Nat. Nanotechnol.* **5**, 497–501 (2010).

283   31.   Luxmoore, I. J. *et al.* Thermal emission from large area chemical vapor deposited graphene

284         devices. *Appl. Phys. Lett.* **103**, 131906 (2013).

285   32.   Lawton, L. M., Mahlmeister, N. H., Luxmoore, I. J. & Nash, G. R. Prospective for graphene

286         based thermal mid-infrared light emitting devices. *AIP Adv.* **4**, 087139 (2014).

287   33.   Nakagawa, K., Takahashi, H., Shimura, Y. & Maki, H. A light emitter based on practicable and

288         mass-producible polycrystalline graphene patterned directly on silicon substrates from a

289         solid-state carbon source. *RSC Adv.* **9**, 37906–37910 (2019).

290   34.   Kim, Y. D. *et al.* Bright visible light emission from graphene. *Nat. Nanotechnol.* **10**, 676–681

291         (2015).

292   35.   Kim, Y. D. *et al.* Ultrafast Graphene Light Emitters. *Nano Lett.* **18**, 934–940 (2018).

293   36.   Barnard, H. R. *et al.* Boron nitride encapsulated graphene infrared emitters. *Appl. Phys. Lett.* **108**,

294         131110 (2016).

295   37.   Berciaud, S. *et al.* Electron and optical phonon temperatures in electrically biased graphene. *Phys.*

296         *Rev. Lett.* **104**, 227401 (2010).

297   38.   Bae, M. H., Ong, Z. Y., Estrada, D. & Pop, E. Imaging, simulation, and electrostatic control of power dissipation in graphene devices. *Nano Lett.* **10**, 4787–4793 (2010).

39. Bae, M. H., Islam, S., Dorgan, V. E. & Pop, E. Scaling of high-field transport and localized heating in graphene transistors. *ACS Nano* **5**, 7936–7944 (2011).

40. Grosse, K. L., Bae, M. H., Lian, F., Pop, E. & King, W. P. Nanoscale Joule heating, Peltier cooling and current crowding at graphene-metal contacts. *Nat. Nanotechnol.* **6**, 287–290 (2011).

41. Mahlmeister, N. H., Lawton, L. M., Luxmoore, I. J. & Nash, G. R. Modulation characteristics of graphene-based thermal emitters. *Appl. Phys. Express* **9**, 012105 (2016).

42. Engel, M. *et al.* Light-matter interaction in a microcavity-controlled graphene transistor. *Nat. Commun.* **3**, 906 (2012).

43. Shi, C., Mahlmeister, N. H., Luxmoore, I. J. & Nash, G. R. Metamaterial-based graphene thermal emitter. *Nano Res.* **11**, 3567–3573 (2018).

44. Luo, F. *et al.* Graphene Thermal Emitter with Enhanced Joule Heating and Localized Light Emission in Air. *ACS Photonics* **6**, 2117–2125 (2019).

45. Planck, M. On the Law of the Energy Distribution in the Normal Spectrum. *Ann. Phys.* **4**, 553–564 (1901).

46. Vora, H., Kumaravadivel, P., Nielsen, B. & Du, X. Bolometric response in graphene based

superconducting tunnel junctions. *Appl. Phys. Lett.* **100**, 153507 (2012).

47. Yan, J. *et al.* Dual-gated bilayer graphene hot-electron bolometer. *Nat. Nanotechnol.* **7**, 472–478 (2012).

48. Mark, J. E. *Physical Properties of Polymers Handbook Second Edition*. *Springer* (2007). doi:10.1007/978-0-387-69002-5_3.

49. Griffiths, P. R., Sloane, H. J. & Hannah, R. W. Interferometers vs Monochromators: Separating the Optical and Digital Advantages. *Appl. Spectrosc.* **31**, 485–495 (1977).

50. Xu, W. & Lee, T. W. Recent progress in fabrication techniques of graphene nanoribbons. *Mater. Horizons* **3**, 186–207 (2016).

51. Pop, E., Varshney, V. & Roy, A. K. Thermal properties of graphene: Fundamentals and applications. *MRS Bull.* **37**, 1273–1281 (2012).

**Acknowledgments**

We thank Ms. S. Sugimoto and Mr. R. Mogi of Keio University, Dr. K. Iwami of Tokyo University of Agriculture and Technology for their technical supports. This work was partially financially supported by Kanagawa Institute of Industrial Science and Technology (KISTEC), PRESTO (Grant Number JPMJPR152B) from JST and KAKENHI (Grant Number 16H04355, 23686055, 18K19025, and 20H02210). This work was technically supported by Spintronics Research Network of Japan, the Core-to-Core program from JSPS, and NIMS Nanofabrication Platform in Nanotechnology Platform Project by MEXT.

**Figures**

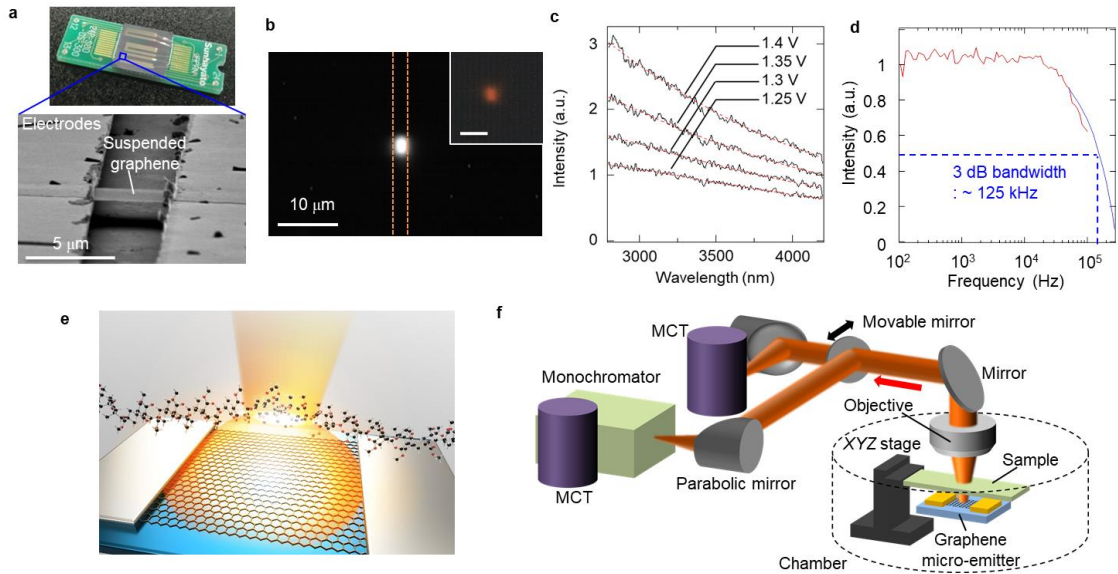

Fig.1 | **Graphene micro-emitters and IR analysis system. a,** A fabricated micro-emitter with suspended graphene on a silicon chip, which is mounted on a chip carrier. **b,** IR and visible (inset) camera images of the light emission from the suspended graphene micro-emitter (5 × 5 μm$^2$) at $V_{ds}$=1.4 V. **c,** Emission spectra from the suspended graphene micro-emitter (5 × 5 μm$^2$) for $V_{p-p}$=1.25–1.4 V with a step of 0.05 V. The red curves are fitted according to Planck's law. **d,** Frequency dependence of the light emission intensity of the suspended graphene micro-emitter (5 × 5 μm$^2$). The red and blue curves show the data obtained by a lock-in amplifier and a spectrum analyzer, respectively. **e,** Schematic of the graphene micro-emitter-based IR analysis. IR light from graphene micro-emitter is irradiated to the specimen, and the transmitted light through the specimen

is measured for IR absorption spectroscopy and imaging. **f,** Schematic of the IR analysis system. IR light generated from graphene micro-emitter is irradiated to the sample, which can be manipulated on a *XYZ* stage. The transmitted light through a specimen is corrected by an objective lens and is guided to the isolated MCT detector or the monochromator with an MCT detector for high-resolution imaging or spectroscopy, respectively

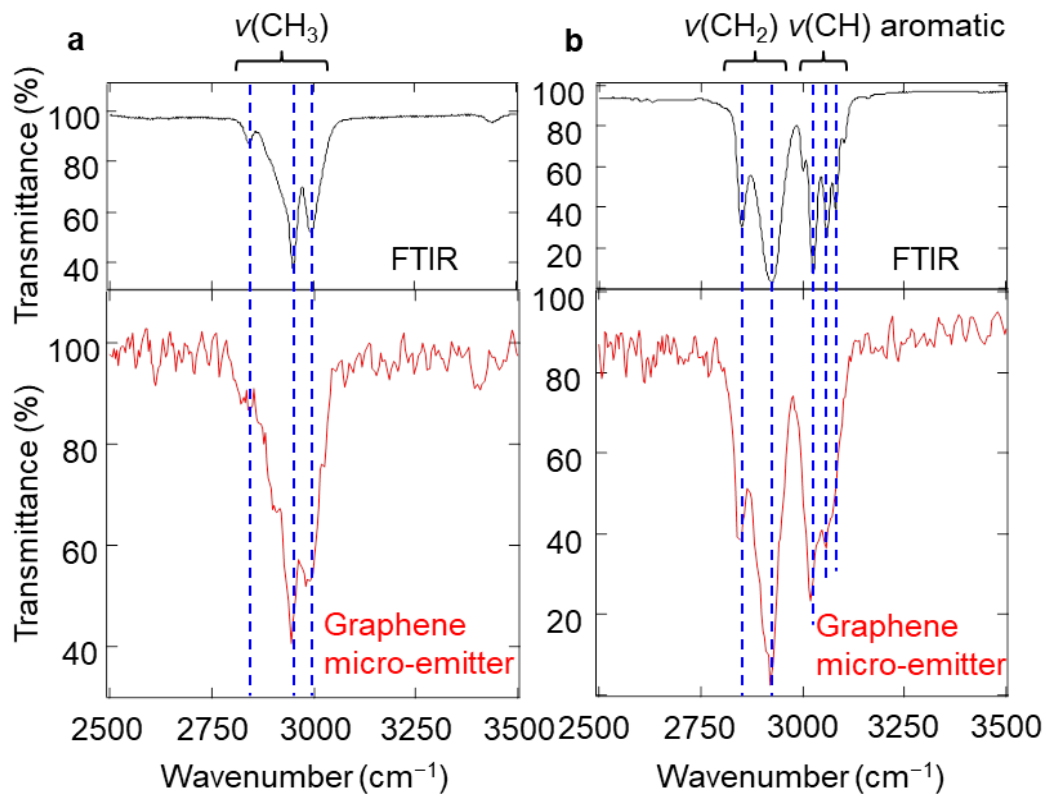

**Fig.2 | IR absorption spectra. a,b,** IR absorption spectra of the polymethyl methacrylate (**a**) and polystyrene (**b**). The upper and lower spectra are the IR absorption spectra measured by conventional FTIR and the IR analysis system by using the suspended graphene micro-emitter (5 × 5 μm$^2$), respectively. The peak positions in the graphene-based IR spectra are coincident to that in the conventional FTIR spectra

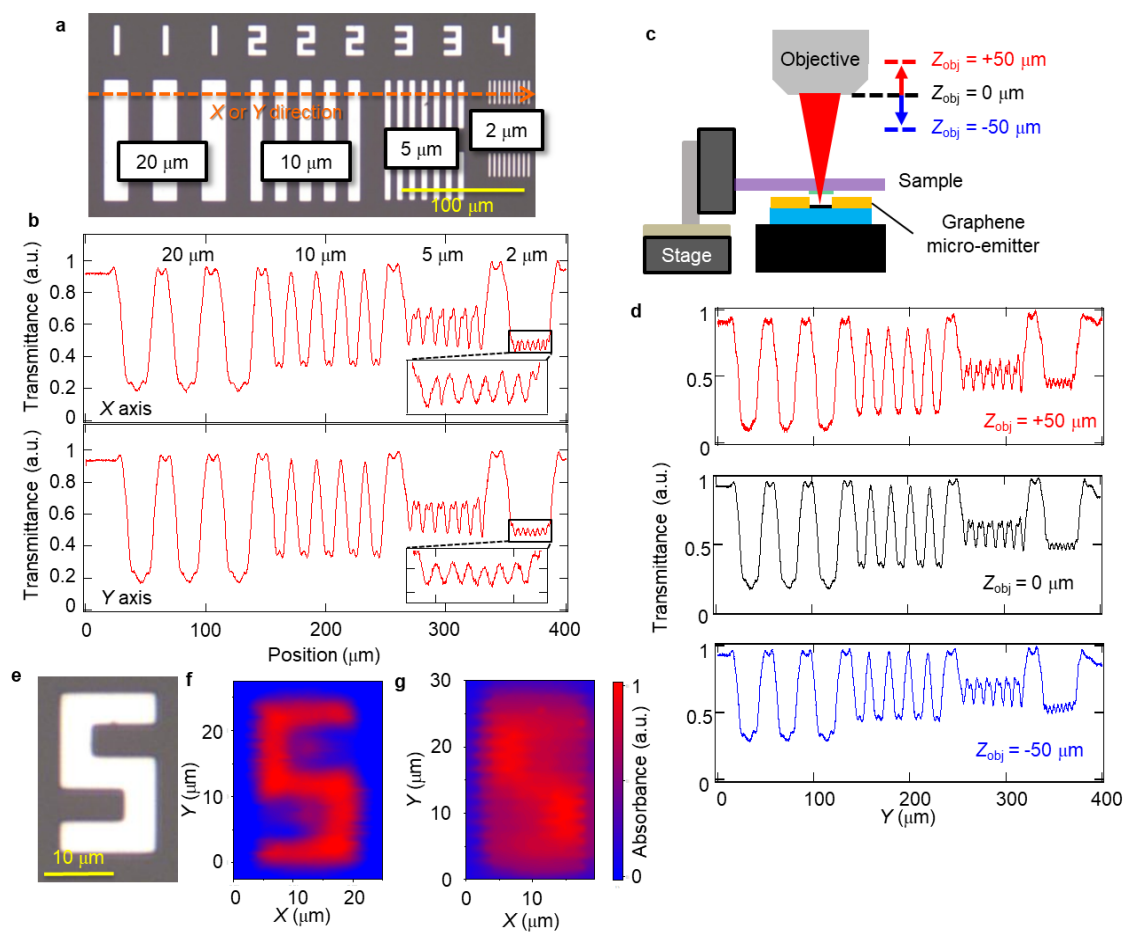

**Fig.3 | High-resolution IR absorption imaging. a,** Optical microscope image of the Ni line and space pattern with the linewidth of 20, 10, 5, and 2 μm. The specimen was moved along the orange arrow, and the transmitted light intensity at each position was measured (one-dimensional line scan). **b,** Transmittance of one-dimensional line scan along the $X$ and $Y$ axes directions by using the unsuspended graphene micro-emitter ($0.5 \times 0.5$ μm$^2$). The one-dimensional line scan was performed by scanning the sample along $X(Y)$ axis with a step of 100 nm. Clear periodical modulation of the transmittance can be observed for all line/space pattern, indicating that spectral resolution is ~2 μm.

**c,** Schematic of the method of the objective position dependence of the spatial resolution. One-dimensional line scan by using the unsuspended graphene micro-emitter (0.5 × 0.5 μm$^2$) was performed while shifting the focus of the objective lens in the ±$z$ direction (|Δ$z$| = 50 μm). **d,** Objective position dependence of the spatial resolution. No line-profile change is observed with changing the position of the objective lens, indicating that spatial resolution does not depend on focusing of the objective lens. **e,** Optical microscope image of the '5'-shaped Ni pattern with the linewidth of 5 μm. **f, g,** Two-dimensional IR absorption imaging of the '5'-shaped Ni pattern in **e** obtained by the unsuspended graphene micro-emitter (0.5 × 0.5 μm$^2$) (**f**) and the conventional FTIR (**g**). The two-dimensional IR absorption imaging by the graphene micro-emitter was performed by scanning the sample in *XY* plane with a step of 500 nm. The spatial resolution of this IR image by the graphene micro-emitter is significantly higher than that by the conventional FTIR microscope

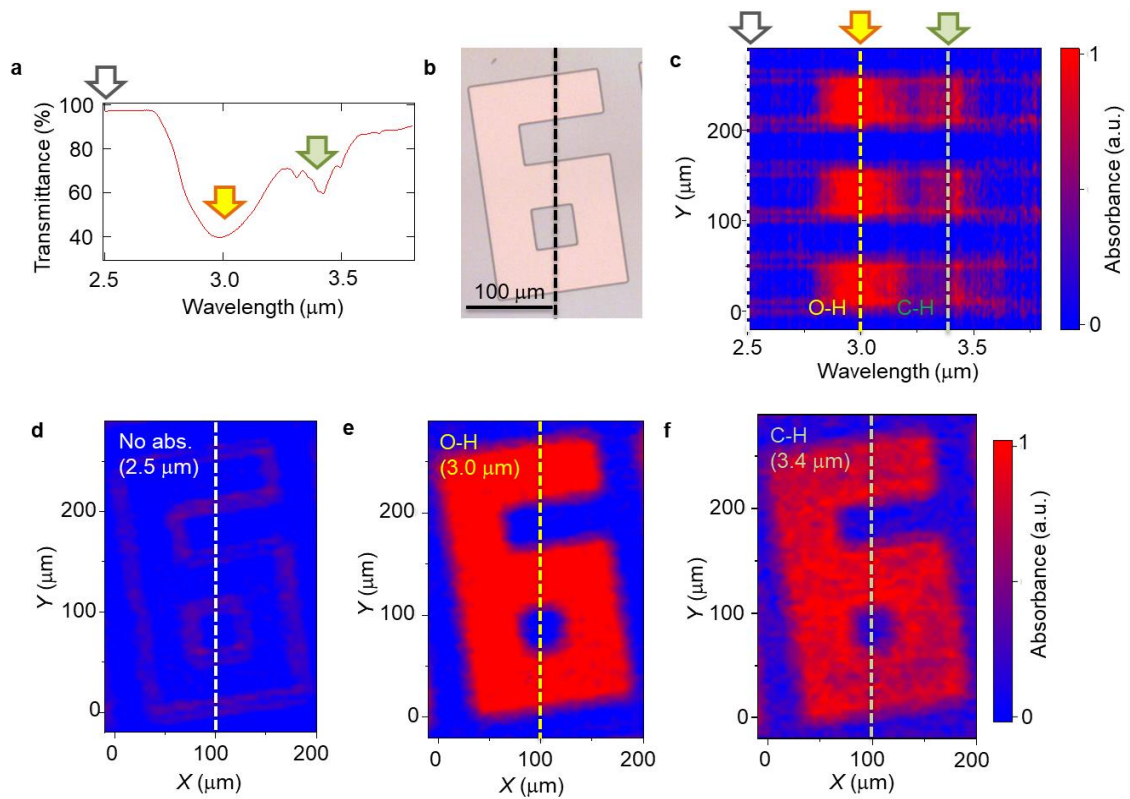

381

382 **Fig. 4 | IR chemical imaging. a,** IR absorption spectrum for photoresist polymer (ZPN1150). Clear

383 dips are observed at 3.0 and 3.4 μm corresponding to the O-H and C-H stretching modes. White,

384 yellow, and green arrows indicate the wavelength of 2.5 μm (no absorption), 3.0 μm (O-H

385 absorption), and 3.4 μm (C-H absorption), respectively. **b,** Optical microscope image of the

386 '6'-shaped ZPN1150 pattern with the linewidth of 50 μm. **c,** Line scan result of absorption spectra

387 along the black broken line on the polymer pattern in **b** by using the unsuspended graphene

388 micro-emitter (10 × 10 μm$^2$). White, yellow, and green arrows and broken lines correspond to the

389 arrows in **a**. The absorption spectra line scan was performed by scanning the sample along *Y* axis with a step of 5 μm. **d, e, f,** Two-dimensional absorption imaging at the wavelength of 2.5 μm (no absorption) (**d**), 3.0 μm (O-H absorption), (**e**) and 3.4 μm (C-H absorption) (**f**), respectively by using the unsuspended graphene micro-emitter ($10 \times 10$ μm$^2$). The two-dimensional absorption imaging was performed by scanning the sample in *XY* plane with a step of 5 μm

## Methods

### Graphene micro-emitter fabrication

Graphene is promising as an IR light source[29,30,31,32,33,34,35,36,37,38,39,40,41,42,43,44] owing to its unique two-dimensional thermal properties.[51] In this study, single side polished Si wafers were used as substrates. $SiO_2$ with a thickness of 1500 nm was deposited on the substrates by chemical vapor deposition. Graphene is mechanically exfoliated from highly ordered pyrolytic graphite and transferred onto the substrates. The graphene on the substrate was patterned by photolithography or electron-beam lithography and $O_2$ plasma etching with a Ni mask, which was subsequently removed by immersion in a HCl solution. Pd (145 nm)/Cr (5 nm) electrodes were deposited on the patterned graphene by vacuum vapor deposition as source and drain electrodes. The sizes of the graphene channels in this work were $5 \times 5$ μm$^2$ for the measurements in Figs. 1 and 2 (Extended Data Fig. 1b), $0.5 \times 0.5$ μm$^2$ for the high-resolution imaging in Fig. 3 (Extended Data Fig. 3), and $10 \times 10$ μm$^2$ for the chemical imaging in Fig. 4 (Extended Data Fig. 1a). For the experiments in Figs. 1 and 2 (Extended Data Fig. 1b), the graphene was suspended by $SiO_2$ etching using vapor HF (Fig. 1a). The electrodes were wire-bonded by Al wires to connect electrically to a chip carrier (Fig. 1a). The electronic property of the fabricated micro-emitters with the suspended graphene of $5 \times 5$ μm$^2$ shows

411 Ohmic conduction behavior, as shown in Extended Data Fig. 1b. Extended Data Fig. 1c shows the

412 frequency dependence of the emission intensity of the graphene micro-emitter with the unsuspended

413 graphene of $5 \times 5$ μm$^2$, in comparison with the conventional high-speed IR emitter (Hawkeye

414 Technologies). The graphene micro-emitter exhibits a stable frequency response over 10 kHz, which

415 is at least 1000 times faster than the conventional IR emitter.

416

417 **Optical measurements**

418 Optical measurements were carried out at room temperature in a high-vacuum chamber

419 (Fig. 1 and Extended Data Fig. 2). In the optical measurement, the emitted light from the graphene

420 micro-emitter was collected by an objective lens through the optical window of the vacuum chamber.

421 The emission image of the graphene micro-emitter under DC bias voltage (Fig. 1b) was directly

422 observed with an InGaAs CCD camera, which has the detection wavelength range from 900 to 1600

423 nm. For the detection of IR light by an MCT detector (detection range, 1–15 μm), which is used in

424 the measurement of emission spectra (Fig. 1c), absorption spectra (Fig. 2), IR imaging (Fig. 3), and a

425 chemical imaging (Fig. 4), graphene emission is modulated by applying a rectangular bias voltage of

426 1113 Hz (1:1 duty ratio) to the graphene micro-emitter by a function generator. In the IR analysis,

the specimen formed on the quartz substrate is mounted to the *XYZ* stage in the vacuum chamber, and the IR light from the graphene micro-emitter is transmitted through the specimen under scanning the sample (Fig. 1f and Extended Data Fig. 2). The transmitted IR light through the specimen is corrected by a Cassegrain objective lens (×40) placed above the vacuum chamber through the sapphire optical window. The collimated light from the objective lens is focused by a parabolic mirror and is guided to an MCT detector directly or through a monochromator with a grating. The intensity of IR light incident to the MCT detector is detected by a lock-in amplifier. The relative spectral response of the optical system such as the optical path and the detector was measured with a standard light from a blackbody furnace, and all spectra were corrected accordingly. For comparison with this novel graphene-based IR analysis system, we measured IR absorption spectra and imaging with the conventional FTIR [ALPHA (Bruker) for the spectra measurements and IRAffinity-1S and AIM-9000 (Shimadzu) for the imaging measurements].

  For the demonstration of the high-resolution IR imaging, we fabricated a sample with the fine pattern of Ni lines and space pattern with the linewidth from 2 to 20 μm (Fig. 3a) and the 5 μm linewidth '5'-shaped Ni pattern (Fig. 3e) on quartz substrates by photolithography technique. These samples were mounted onto a *XYZ* piezo motor stage in the vacuum chamber. The sample was brought into close contact with the graphene micro-emitters, and IR transmittance (absorption) were obtained by scanning the sample along $X(Y)$ axis with a step of 100 nm for one-dimensional line scan and in $XY$ plane with a step of 500 nm for two-dimensional IR absorption imaging. Since a mechanically stable and small-footprint graphene micro-emitter is necessary for the high resolution imaging, we used an unsuspended graphene emitter with the sub-micrometer size of $0.5 \times 0.5$ μm$^2$. As described in the main text and Fig. 3, the graphene-based IR analysis method has high spatial resolution of ~2 μm. As shown in the line scan results of the conventional FTIR in Extended Data Fig. 3, the spatial resolution of this method is significantly higher than that of the conventional FTIR microscopy.

For the demonstration of the IR chemical imaging, we fabricated the 50-μm width '6'-shaped polymer pattern of ZPN1150 photoresist (Zeon Corporation) on a quartz substrate (Fig. 4b). This polymer has absorption peaks of O-H and C-H stretching modes at 3.0 and 3.4 μm, respectively (Fig. 4a). Since a mechanically stable and bright graphene micro-emitter is necessary for the simultaneous measurement of spectroscopy and imaging though a monochromator, we used an unsuspended graphene micro-emitter with the large footprint of $10 \times 10$ μm$^2$ (Extended Data Fig. 1a). We measured the line scan of absorption spectra across the '6'-shaped polymer pattern along $Y$

459 axis with a step of 5 μm, where the absorption spectra are measured at each position. We also

460 measured the two-dimensional absorption imaging at the wavelength of 2.5 μm (no absorption), 3.0

461 μm (O-H absorption), and 3.4 μm (C-H absorption). In this imaging, the measured wavelength of the

462 grating in a monochromator is set to their wavelength, and the IR intensity though the sample is

463 measured under scanning in *XY* plane with the step of 5 μm.

464

465 **Extended Data**

466 **High-resolution IR spectroscopy and imaging**
467 **based on graphene micro-emitters**

468

469 Kenta Nakagawa,[1,2,‡] Yui Shimura,[1,‡] Yusuke Fukazawa,[1] Ryosuke Nishizaki,[1] Shinichiro Matano,[1]

470 and Hideyuki Maki[1,3,*]

471

472 [1]Department of Applied Physics and Physico-Informatics, Keio University, 3-14-1 Hiyoshi,

473 Kohoku-ku, Yokohama, Kanagawa 223-8522, Japan.

474 [2]Kanagawa Institute of Industrial Science and Technology (KISTEC), 705-1 Shimoimaizumi, Ebina,

475 Kanagawa 243-0435, Japan.

476 [3]Center for Spintronics Research Network, Keio University, 3-14-1 Hiyoshi, Kohoku-ku, Yokohama,

477 Kanagawa 223-8522, Japan.

478 ‡These authors contributed equally to this work.

479 *E-mail: maki@appi.keio.ac.jp

480

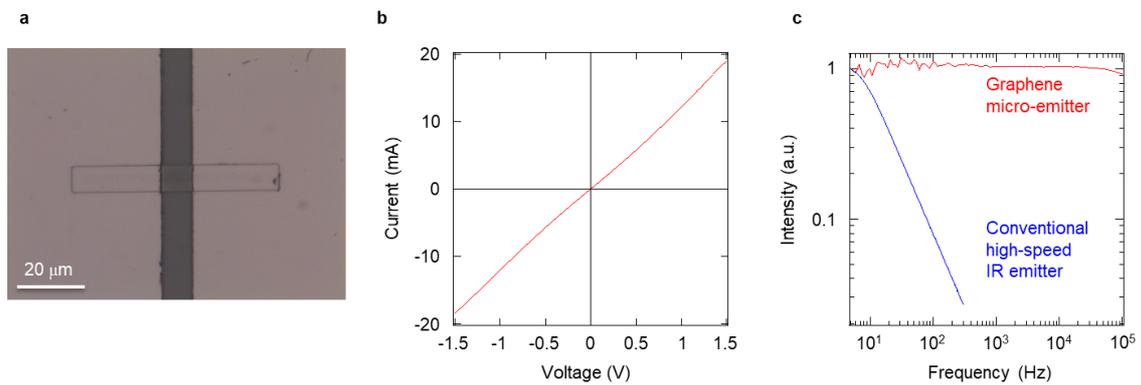

**Extended Data Fig. 1 | Graphene micro-emitter. a,** Optical microscope image of the unsuspended graphene micro-emitter with the size of 10 × 10 μm$^2$. **b,** DC bias voltage ($V_{ds}$) dependence of the current $I$ ($I$-$V_{ds}$ curve) of the suspended graphene micro-emitter with the size of 5 × 5 μm$^2$. The fabricated micro-emitters show Ohmic conduction behavior. **c,** Frequency dependence of the emission intensity of the unsuspended graphene micro-emitter with the size of 5 × 5 μm$^2$, in comparison with the conventional high-speed IR emitter (Hawkeye Technologies). The graphene micro-emitter exhibits stable frequency response over 10 kHz, which is at least 1000 times faster than the conventional IR emitter

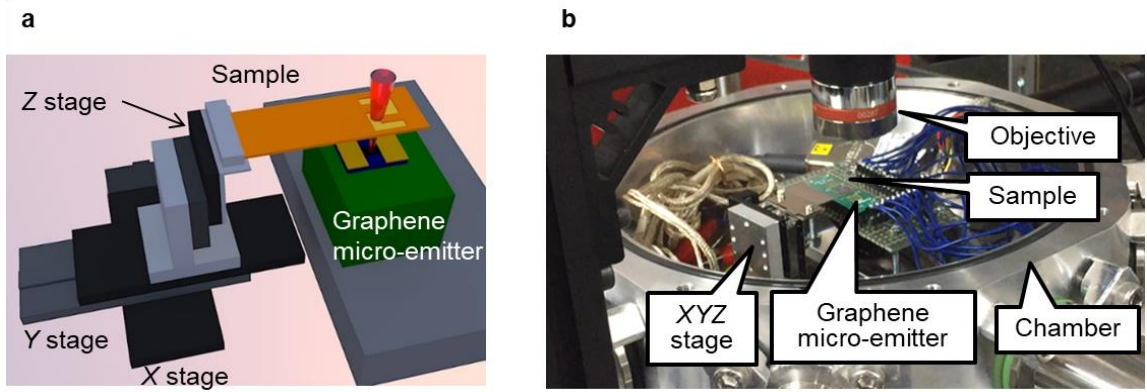

**Extended Data Fig. 2 | System in the IR system vacuum chamber. a,b,** Schematic image (**a**) and photograph (**b**) of the system in the IR analysis system vacuum chamber. In the analysis vacuum chamber, the specimen is mounted to the *XYZ* stage, and the IR light from the graphene micro-emitter is transmitted through the sample. The transmitted IR light through the specimen is corrected by a Cassegrain objective (×40) placed above the vacuum chamber through the sapphire optical window. For the measurement of the IR imaging, the specimen is mounted to the *XYZ* piezo motor stage. The specimen is brought into close contact to the graphene micro-emitter, and IR transmittance (absorption) images are obtained by scanning the sample in *XY* plane

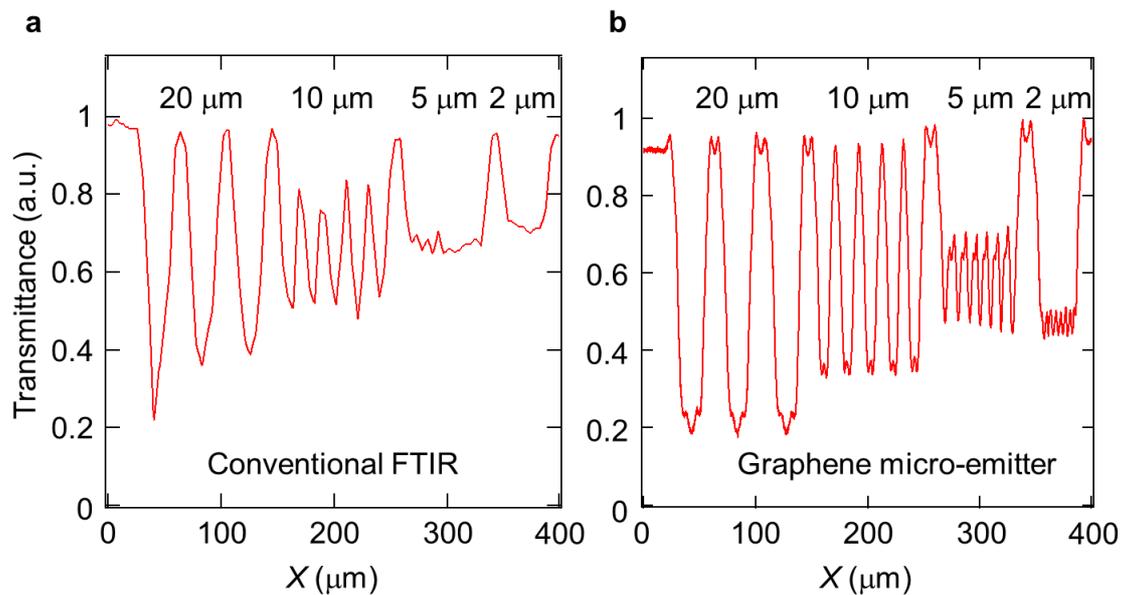

**Extended Data Fig. 3 | Comparison of the spatial dispersion between conventional FTIR and graphene micro-emitter. a,b,** Transmittance of one-dimensional line scan along the *X* axis direction by using conventional FTIR (**a**) and the unsuspended graphene micro-emitter (0.5 × 0.5 μm$^2$) (**b**). Clear periodical modulation of the transmittance can be observed for all line/space pattern by using the graphene micro-emitter in contrast to the conventional FTIR